# Mesoscopic Superconducting Cylinders:

# Phase Transitions, Phase Diagrams and Magnetization

W. V. Pogosov
Moscow Institute of Physics and Technology, 141700 Dolgoprudnyi, Moscow Region, Russia

A. L. Rakhmanov
Institute for Theoretical and Applied Electrodynamics RAS, 127412 Moscow, Russia

E. A. Shapoval
All-Russia Scientific-Research Institute of the Metrological Service, Moscow, 119361 Russia

*Abstract*—Superconducting state in mesoscopic cylinder placed in the external magnetic field is analyzed at general de Gennes boundary condition for the order parameter. The lower and the surface critical fields of the cylinder are calculated at different values of de Gennes "extrapolation length" *b*. For the case of standard boundary condition (corresponding to $b \to \infty$) we use a variational method to calculate the phase transitions between different superconducting states. Based on the trial function for the order parameter, we find an approximate solutions to the Ginzburg-Landau equations for the Meissner, the single-vortex, and the giant-vortex states. The magnetization curves of thin cylinders are calculated.

## I. Introduction

The vortex phases in mesoscopic superconductors are under considerable current interest now (see, for example [1]-[6]). It is known that a type-II superconductor is characterized by the lower $H_{c1}$, upper $H_{c2}$, and surface $H_{c3}$ critical fields. Usually, the phase diagram of the superconductor has the following main features. Below $H_{c1}$ the vortex-free state (the Meissner state) is the most energetically favorable phase. The multi-vortex state (the Abrikosov phase) becomes energetically more favorable between $H_{c1}$ and $H_{c2}$. In the vicinity of $H_{c2}$ the phase with the giant-vortex in the center of the sample carrying $L > 1$ flux quanta has the lowest energy, and Abrikosov vortices merge into this structure. In this case, the order parameter is strongly suppressed in the inner part of the sample, and this state can be referred to a surface superconductivity. With further increasing of the applied magnetic field $H$, phase transitions between different giant-vortex states are observed, and the superconductor comes to the normal state at $H = H_{c3}$. Note that both the values of $H_{c1}$ and $H_{c3}$ depend on the sample sizes and geometry.

In this paper, we study the phase diagram of type-II superconducting cylinder allowing for general de Gennes boundary condition for the order parameter *f* [9]:

$$f'(R) + b^{-1} f(R) = 0, \qquad (1)$$

This work is supported by the Russian Foundation for Basic Research (RFBR), grants nos. 00-02-18032, 00-15-96570, and 01-02-06526, by the joint INTAS-RFBR program, grant no. IR-97-1394, and by the Russian State Programs 'Integracia' and 'Fundamental Problems in Condensed Matter Physics'.

where *b* is "extrapolation length", *R* is the cylinder radius. This condition should be used for the interface of superconductor and normal metal, for anisotropic and high-$T_c$ superconductors, or when the surface of the superconductor has some defects. Here and below we use a dimensionless variables: the distances, the magnetic field *H*, and the order parameter are measured in units of $\xi(T)$, $H_{c2}$, and $\sqrt{-\alpha/\beta}$, respectively, where $\xi(T)$ is the coherence length, $\alpha$, and $\beta$ are the Ginzburg-Landau (GL) coefficients. First, we calculate the dependences of $H_{c1}$ and $H_{c3}$ on the cylinder radius at different values of *b* at GL parameter $\kappa >> 1$. This allows us to reveal the main features of the phase diagram of the cylinder at different *b*. We show that the phase diagram depends appreciably on value of *b*. After this, we study in a more detail the phase diagram of the cylinder at standard boundary condition ($b \to \infty$) and arbitrary $\kappa$. For this purpose, we apply a variational approach. Based on the trial function for the order parameter, we solve GL equations for the Meissner state, single-vortex state, and giant-vortex state. The model enables us to find explicit expressions for the dependence of the lower critical field of the cylinder $H_{c1}$ on its radius and $\kappa$. Phase boundaries between different superconducting states are studied and an equilibrium phase diagram of thin cylinder is obtained. We calculate the magnetization curve of thin cylinder, which can carry only several quanta of magnetic flux.

## II. Critical fields at different *b*

If the order parameter has the axial symmetry inside long cylinder, it can be presented as $f(r)\exp(i\varphi L)$, where *r* is the dimensionless radius-vector in the cross-sectional area of the cylinder, $\varphi$ is the azimuthal angle, and *L* is the angular quantum momentum ($L = 0, 1, 2,\ldots$). If $L = 0$ this state is the Meissner state, if $L = 1$ it is the single-vortex state, and at $L > 1$ it is the giant-vortex state.

In the vicinity of $H_{c3}$ the magnetic field is assumed to be constant throughout the cylinder and the first GL equation can be linearized. The solution of the resulting equation is [10]:

$$f(r, L) = r^L \exp\left(\frac{Hr^2}{4}\right) \Phi\left(\frac{H-1}{2H}, L+1, \frac{Hr^2}{2}\right), \qquad (2)$$

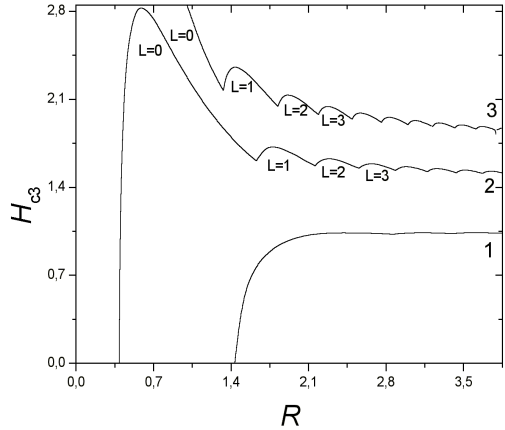

**Fig. 1.** The surface critical field versus cylinder radius $R$ at different values of the Gennes "extrapolation length" $b$: $b = 1$ (curve 1), $b = 5$ (curve 2), $b \to \infty$ (curve 3). The numbers below curves denote the angular quantum momentum of superconducting states. $H_{c3}$ and $R$ are measured in units of the upper critical field $H_{c2}$ and the coherence length $\xi(T)$, respectively.

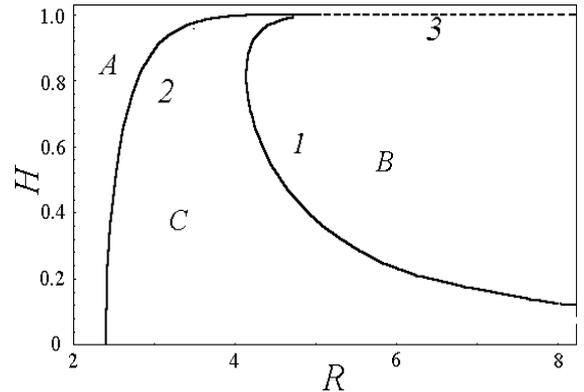

**Fig. 2.** The equilibrium phase diagram of the cylinder at $b = 0$ and $\kappa \gg 1$. Region A - normal state, B – vortex state, C – superconducting vortex-free state. Curve 1 shows the lower critical field, curve 2 shows the boundary between the normal state and the vortex-free state, curve 3 shows the surface critical field. $H$ and $R$ are measured in units of the upper critical field $H_{c2}$ and the coherence length $\xi(T)$, respectively.

where $\Phi$ is Cummer function. Expression (2) should meet boundary condition (1) at the cylinder surface, therefore, $H_{c3}$ is determined by $b$ and $R$. At each $R$ one should choose the maximum value of $H_{c3}$ among those, corresponding to different $L$. The resulting dependence $H_{c3}(R)$ is shown in Fig. 1 at $b = 1$ (curve 1), $b = 5$ (curve 2), and $b \to \infty$ (curve 3). The numbers below the curves $H_{c3}(R)$ characterize the angular quantum momentum $L$ of corresponding vortex states. It is clearly seen that the behavior of $H_{c3}(R)$ depends appreciably on $b$. The surface superconductivity is suppressed with decreasing of $b$. At $b = 1$ and $R \to \infty$ the difference between $H_{c3}(R)$ and $H_{c2}$ is just a few percent, and the oscillations of $H_{c3}(R)$ are practically indistinguishable on curve 1.

The lower critical field is the value of the external magnetic field at which the energies of the Meissner state and the single-vortex state become equal. In both cases the order parameter and the magnetic field have axial symmetric distributions inside the cylinder. When $\kappa \gg 1$, it can be supposed that the magnetic field is constant throughout the sample. In this case, in order to find the energies of the Meissner and the single-vortex states one has to solve a single nonlinear first GL equation. We have solved this equation for two limiting cases of the boundary condition for the order parameter, $b = 0$ and $b \to \infty$. The former case corresponds to the interface of superconductor and normal metal, the latter to the interface of low-$T_c$ superconductor and vacuum. The resulting dependences $H_{c1}(R)$ are shown in Fig. 2 and 3 (curves 1) at $b = 0$ and $b \to \infty$, respectively. In the same figures the dependences $H_{c3}(R)$ are also plotted. It is seen from the phase diagram in Fig.2 that the giant vortex phases do not nucleate at $b = 0$.

Let us discuss the phase diagram at $b = 0$ (Fig. 2). It includes the normal state (region A), the vortex state (region B), and the vortex-free superconducting state (region C). In decreasing applied field, when $R > 4.14$ the cylinder first passes from the normal state to the superconducting vortex-free state (to the region C rapidly narrowing with increasing $R$). Then at $H = H_{c1}(R)$ (upper branch of curve 1) it passes to the region of the vortex state $B$. Finally, at $H = H_{c1}(R)$ (lower branch of curve 1) it passes again to the vortex-free state C. When $R < 4.14$ the cylinder comes from the normal to the Meissner state by-passing the vortex phase (curve 2). The surface critical field is smaller than $H_{c2}$ (curve 3) at $b = 0$. The phase diagram at $b \to \infty$ is studied in a more detail in the next section.

### III. PHASE TRANSITIONS AT STANDARD BOUNDARY CONDITION

In this section we study the phase transitions between different superconducting states with axial symmetric distributions of the order parameter and the magnetic field at standard boundary condition ($b \to \infty$). For this purpose, we use a variational method and solve GL equations using trial function for the order parameter:

$$f = \begin{cases} f_L \left(\dfrac{r}{r_L}\right)^L \left(2 - \dfrac{r^2}{r_L^2}\right)^{L/2}, & r \leq r_L, \\ f_L, & r \geq r_L, \end{cases} \quad (3)$$

where $f_L$ and $r_L$ are variational parameters. The values of $f_L$ and $r_L$ can be calculated self-consistently by minimization of the total GL free energy of the cylinder. Comparing the energies of the states with different $L$ one can calculate the phase boundaries corresponding to the transitions from $L$ to $L + 1$. The resulting equilibrium phase diagram is shown in Fig.3 at $\kappa = 5$. We found that the function $H_{c1}(R)$ behaves in a similar way (scaling behavior) starting with $\kappa \approx 2$. Besides, at large $\kappa$ the dependences $H_{c1}(R)$ calculated by the variational and the numerical methods (see the previous section) are practically indistinguishable from each other (curve 1, Fig. 3). In Fig. 4 for illustration we plotted the typical distribution of the order parameter inside the cylinder calculated within our variational approach. This state, which is most favorable energetically, represents the giant vortex phase with $L = 3$.

The energy of the superconductor in the Abrikosov phase, which can be energetically more favorable below

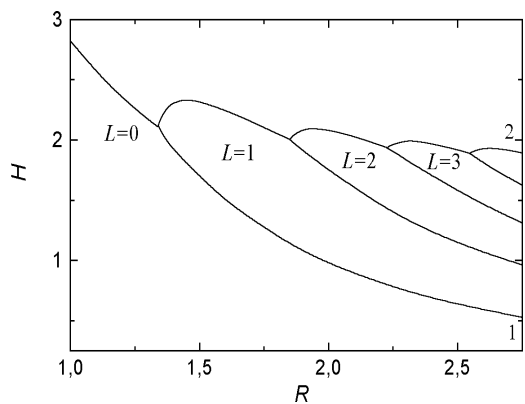

**Fig. 3.** The equilibrium phase diagram of the cylindrical type-II superconductor ($\kappa = 5$, $b \to \infty$) in the applied magnetic field calcularred by variational method. Curve 1 corresponds to the lower critical field $H_{c1}$; curve 2 corresponds to the surface critical field $H_{c3}$ calculated by the variational method. The numbers below curve 2 denote the angular quantum momentum of superconducting states. $H$ and $R$ are measured in units of the upper critical field $H_{c2}$ and the coherence length $\xi(T)$, respectively.

$H = H_{c2}$, can not be calculated in our model since the order parameter is not axially symmetric in this case. Nevertheless, as can be seen from the phase diagram (Fig. 3), the axial-symmetric states have the lower energy as compared to the Abrikosov phase at small cylinder radius $R \leq 2.75$, since the field of transition from $L = 1$ to $L = 2$ state is bigger than $H = H_{c2}$ in this case.

The model enables us to calculate the magnetization curves of the cylinders with different radiuses. The typical dependence is shown in Fig. 5 at $R = 2.5$. In this case the cylinder is able to accommodate the giant vortex with the maximum value of $L$ equal to 3. Jumps in the magnetization correspond to the transitions between different vortex phases.

## IV. CONCLUSION

We calculated the dependences of the lower and the surface critical fields of long superconducting cylinder at different cylinder radiuses and at different boundary conditions for the order parameter. This allows us to reveal the main features of the phase diagram of the cylinder. We showed that the phase diagram depends appreciably on the boundary condition for the order parameter. We proposed a variational model to study different vortex phases at standard boundary condition at the cylinder surface and calculated its phase diagram and magnetization.

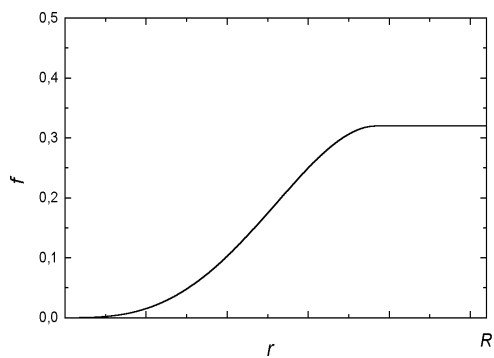

**Fig. 4.** The spatial distribution of the order parameter in the cylinder at $\kappa = 5$, $b \to \infty$, $H = 1.75\, H_{c2}$, $R = 2.6$, $r$ is the distance from the cylinder axis. The giant vortex with angular quantum momentum $L = 3$ is located at the cylinder axis.

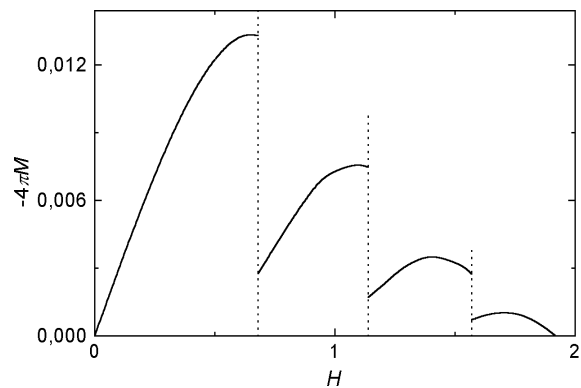

**Fig. 5.** The equilibrium field dependence of the magnetization of the thin cylinder with the radius $R = 2.5$ at $\kappa = 5$, $b \to \infty$. Jumps in the magnetization correspond to the transitions between different $L$ - states. $M$ and $R$ are measured in units of the upper critical field $H_{c2}$ and the coherence length $\xi(T)$, respectively.

## ACKNOWLEDGMENTS

We are grateful to K. I. Kugel, L. G. Mamsurova, and K. S. Pigalskiy for useful discussions.